\documentclass{aastex61}

 \usepackage{CJK}
\begin{document}

 \begin{CJK*}{UTF8}{gbsn}

\title{The hidden magnetic structures of a solar intermediate filament revealed by the injected flare material}

\correspondingauthor{Xiaoli Yan}
\email{yanxl@ynao.ac.cn}

\author{Xiaoli Yan}
\affiliation{Yunnan Observatories, Chinese Academy of Sciences, Kunming 650216, People's Republic of China.}
\affiliation{Yunnan Key Laboratory of Solar Physics and Space Science, Kunming, 650216, People's Republic of China.}

\author{Zhike Xue}
\affiliation{Yunnan Observatories, Chinese Academy of Sciences, Kunming 650216, People's Republic of China.}
\affiliation{Yunnan Key Laboratory of Solar Physics and Space Science, Kunming, 650216, People's Republic of China.}

\author{Jincheng Wang}
\affiliation{Yunnan Observatories, Chinese Academy of Sciences, Kunming 650216, People's Republic of China.}
\affiliation{Yunnan Key Laboratory of Solar Physics and Space Science, Kunming, 650216, People's Republic of China.}

\author{Liheng Yang}
\affiliation{Yunnan Observatories, Chinese Academy of Sciences, Kunming 650216, People's Republic of China.}
\affiliation{Yunnan Key Laboratory of Solar Physics and Space Science, Kunming, 650216, People's Republic of China.}

\author{Kaifan Ji}
\affiliation{Yunnan Observatories, Chinese Academy of Sciences, Kunming 650216, People's Republic of China.}

\author{Defang Kong}
\affiliation{Yunnan Observatories, Chinese Academy of Sciences, Kunming 650216, People's Republic of China.}
\affiliation{Yunnan Key Laboratory of Solar Physics and Space Science, Kunming, 650216, People's Republic of China.}

\author{Zhe Xu}
\affiliation{Yunnan Observatories, Chinese Academy of Sciences, Kunming 650216, People's Republic of China.}
\affiliation{Yunnan Key Laboratory of Solar Physics and Space Science, Kunming, 650216, People's Republic of China.}

\author{Qiaoling Li}
\affiliation{School of Physics and Astronomy, Yunnan University, Kunming, 650050, Peopleʼs Republic of China.}

\author{Liping Yang}
\affiliation{Yunnan Observatories, Chinese Academy of Sciences, Kunming 650216, People's Republic of China.}

\author{Xinsheng Zhang}
\affiliation{Yunnan Observatories, Chinese Academy of Sciences, Kunming 650216, People's Republic of China.}

\begin{abstract}
Solar filaments are spectacular objects in the solar atmosphere, consisting of accumulations of cool, dense, and partially ionized plasma suspended in the hot solar corona against gravity. The magnetic structures that support the filament material remain elusive, partly due to the lack of high resolution magnetic field measurements in the chromosphere and corona. In this study, we reconstruct the magnetic structures of a solar intermediate filament using EUV observations and two different methods, to follow the injection of hot material from a B-class solar flare. Our analysis reveals the fine-scale magnetic structures of the filament, including a compact set of mutually wrapped magnetic fields encasing the cool filament material, two groups of helical magnetic structures intertwining with the main filament, and a series of arched magnetic loops positioned along the filament. Additionally, we also find that the northern footpoints of the helical structures are rooted in the same location, while their southern footpoints are rooted in different areas. The results obtained in this study offer new insights into the formation and eruption mechanisms of solar filaments.
\end{abstract}
 

\section{Introduction}\label{sec:introduction}
Solar filaments, also known as prominences, are considered the same phenomena whether observed on the solar disk or above the solar limb. They are believed to be embedded in the coronal magnetic field, characterized by predominantly horizontal and dipped magnetic structures (Tandberg-Hanssen, 1995). Eruptions of solar filaments are frequently linked to coronal mass ejections (CMEs), which are among the largest explosive events in the solar system (Zhang \& Wang, 2001; Chen, 2011; Yan et al., 2011; Zhang et al., 2015; Song et al., 2019; Liu, 2020). Researchers studying the morphology of CMEs suggest that solar filaments serve as the cores of these events (Forbes, 2000).

Magnetic field observations in the photosphere indicate that solar filaments form along the neutral line between regions of opposite polarity (Babcock \& Babcock, 1955; Martin, 1998). Based on their formation locations, solar filaments are classified into three categories: quiescent filaments, active-region filaments, and intermediate filaments (Engvold, 1998). Typically, quiescent and intermediate filaments are found to develop within filament channels (Gaizauskas, 1998).

Observations and simulations suggest that solar filaments possess two distinct magnetic structures (see recent reviews by Mackay et al., 2010; Parenti, 2014; Vial \& Engvold, 2015, and references therein). One structure is the sheared arcade (Kippenhahn \& Schluter, 1957; Antiochos et al., 1994), while the other is the twisted flux rope (Kuperus \& Raadu, 1974; van Ballegooijen \& Martens, 1989; Aulanier \& Demoulin, 1998). However, distinguishing between these two magnetic structures is challenging due to the limitations of two-dimensional imaging and the lack of magnetic field measurements in the chromosphere and corona. Based on Martin's (1998) work, Chen et al. (2014) developed a new method to differentiate between the two structures based on the orientations of barbs/threads and the overlying magnetic loops. Nevertheless, this approach remains indirect and is constrained by the reliance on two-dimensional observations. Currently, there is limited observational evidence supporting the sheared arcade structure, with increasing observations favoring the twisted magnetic structure. For instance, twisting and untwisting motions are frequently observed before and during solar filament eruptions (Schmieder et al., 1985; Yan et al., 2014a,b; Yang et al., 2014; Xue et al., 2016; Li et al. 2021). Magnetic flux ropes are often identified at the locations of solar filaments through non-linear force-free extrapolation (Guo et al., 2010; Jiang et al., 2014; Wang et al., 2015; Yan et al., 2015, 2016). However, these findings also belong to indirect evidence. While some studies have observed signs of twisted magnetic structures during the activation and eruption phases of filaments (Wang et al., 1996; Filippov et al., 2015; Wang \& Liu, 2019; Okamoto et al., 2016), it remains difficult to determine when these twisted structures form. Some observations indicate that twisted magnetic structures exist before filament eruptions (Wang et al., 2015; Filippov et al., 2015), while others suggest they form during the eruption itself (Yan et al., 2020a).

Previous investigations using spectrum inversions have shown that prominences at the solar limb primarily consist of horizontal magnetic fields relative to the solar surface (Casini et al., 2003; Lopez Ariste \& Aulanier, 2007; Schmieder et al., 2014). Bak-Steslicka et al. (2013) measured linear polarization and line-of-sight velocity in the prominence cavity at the solar limb using the Coronal Multi-Channel Polarimeter (CoMP). Their findings revealed a bulls-eye pattern in Doppler velocity and linear polarization signatures within the prominence cavities, which support the idea that the magnetic field structure of prominences resembles a flux rope (Gibson \& Fan, 2006; Zhang et al., 2012; Cheng et al., 2017). Consequently, the question of the magnetic structures of solar filaments remains controversial.

With the advent of ground-based and space-borne solar telescopes, high-resolution observations have revealed that filaments are composed of numerous fine threads oriented at angles to their spine (Lin et al., 2007; Yan et al., 2015; Shen et al., 2015; Li et al., 2018). Additionally, counter-streaming flows have been observed along the filament spine and barbs (Zirker et al., 1998). Complex material movement and magnetic convection have also been documented in limb prominences (Berger et al., 2010, 2011).

Fortunately, data from the Solar Dynamics Observatory (SDO) and the New Vacuum Solar Telescope (NVST) captured the fine magnetic structures of a solar intermediate filament before its eruption, tracing the injection of hot material from a B-class flare near one end of the filament's footpoints. The paper is organized as follows: The observations and methods are described in Section 2. The results are presented in Section 3. The discussion are given in Section 4.

\section{Observations and methods}\label{sec:observations}
\subsection{Observations}
The New Vacuum Solar Telescope (NVST) is a ground-based telescope with a 986 mm clear aperture located at the Fuxian Solar Observatory, part of the Yunnan Observatories, Chinese Academy of Sciences (CAS) (Liu et al., 2014; Yan et al., 2020b). It is designed to observe the fine structures of the Sun’s photosphere and chromosphere with high spatial resolution ($0.04^{\prime\prime}$ pixel$^{-1}$ for TiO images and $0.165^{\prime\prime}$ pixel$^{-1}$ for H$\alpha$ images) and high temporal resolution (approximately 12 seconds), making it well-suited for investigating small-scale solar activities (Tian et al., 2016). The H$\alpha$ line-center and off-band channels ($\pm$0.5 \AA) at 6562.8 ~{\AA}  are used to show filament evolution and construct Doppler maps. The calibrated NVST images are reconstructed using the speckle masking method (Xiang et al., 2016), and the coalignment of the NVST H$\alpha$ images follows the method of Cai et al. (2022). Additionally, H$\alpha$ images and off-band channels from the Optical and Near-infrared Solar Eruption Tracer (ONSET) are employed to illustrate the flare material injection and construct Doppler maps (Fang et al., 2013). The ONSET H$\alpha$ images have a pixel size of $0.4^{\prime\prime}$ and a time cadence of 30 seconds. Full-disk H$\alpha$ images from the Global Oscillation Network Group (GONG) are also utilized to show material injection and the B-class flare (Harvey et al., 1996), with a spatial resolution of $1^{\prime\prime}$ pixel$^{-1}$ and a time cadence of 1 minute.

EUV images obtained by the Atmospheric Imaging Assembly (AIA) on board the Solar Dynamics Observatory (SDO) (Lemen et al., 2012; Pesnell et al., 2012) include channels at 304~\AA, 171~\AA, 193~\AA, 335~\AA, 211~\AA, 94~\AA, and 131~\AA. These images capture the evolution of the filament during the flare material injection, and six of the EUV wavelengths are used to derive the differential emission measure (DEM). The spatial resolution of these data is $1.^{\prime\prime}5$s, with a time cadence of 12 seconds. Additionally, photospheric line-of-sight (LOS) magnetograms provided by the Helioseismic and Magnetic Imager (HMI) on board SDO (Schou et al., 2012) have a cadence of 45 seconds and a pixel scale of 0.$^\prime$$^\prime$5. These magnetograms are used to illustrate the magnetic fields surrounding the filament location.

EUV images at 304 \AA\ and 195 \AA\ from the Extreme Ultra-Violet Imager (EUVI) on board the Solar Terrestrial Relations Observatory Ahead (STEREO-A) (Wuelser et al., 2004; Kaiser et al., 2008; Howard et al., 2008) are used to display the morphology of the filament at the solar limb. The 304 \AA\ images have a spatial resolution of $1.^{\prime\prime}6$ $pixel^{-1}$ and a cadence of 10 minutes, while the 195 \AA\ images have a time cadence of approximately 3 minutes.

\subsection{Methods}
The Dopplergrams are constructed by using the following equation (Langangen et al. 2008), 
\begin{equation}
D = \frac{B - R}{B + R},
\end{equation}
where B and R represent the H$\alpha$ blue-wing (-0.5 \AA) and red-wing (+ 0.5 \AA) images, respectively. Note that the Dopplergrams just show the Doppler shift signals (not the exact Doppler velocity).

We calculate the differential emission measure (DEM) using the almost simultaneous observations of six AIA EUV lines (131 \AA, 94 \AA, 335 \AA, 211 \AA, 193 \AA, and 171 \AA\ formed at coronal temperatures). The DEM is determined by 
\begin{equation}
I_{i} = \int R_{i}(T) \times DEM(T)dT,
\end{equation}
where $I_{i}$ is the observed intensity of the waveband $i$,  $R_{i}(T)$ represents the temperature response function of waveband $i$, and $DEM(T)$ is the DEM of coronal plasma at temperature T, which is computed using the routine \texttt{xrt\_dem\_iterative2.pro} in the Solar Software package. This code was first written for Hinode/X-ray Telescope data (Golub et al. 2007; Weber et al. 2004), and then modified for SDO/AIA data (Cheng et al. 2012). In this study, log $T$ is set in the range of 5.7--7.3, where the DEM is generally well constrained (Aschwanden \& Boerner 2011; Hannah \& Kontar 2012; Cheung et al. 2015; Plowman \& Caspi 2020). 

To obtain the emission measure (EM) in the temperature ranges $[T_\mathrm{min},\,T_\mathrm{max}]$, we evaluate
\begin{equation}
EM=\int _{T_{min}}^{{T}_{max}}DEM(T)dT. 
\end{equation}

The DEM-weighted average temperature is calculated as follows:
\begin{equation}
\bar{T}=\frac{\int DEM(T) \times TdT}{\int DEM(T)dT}. 
\end{equation}

The errors of the DEM solutions are very small in the temperature range of 5.7 $\leq$ $\log$ T $\leq$ 7.3 , indicating that the DEM is well constrained by the AIA data over most of the temperature range of the flare region. 

The maximum fusion method employed in this paper involves extracting the intensity values at each pixel from four images taken at different moments. The maximum value for each pixel across these four images is then selected to reconstruct a single image. This approach highlights the complete structure of the filament by utilizing images captured at different times.

\section{Results}\label{sec:results}
\subsection{Appearance of the solar intermediate filament}
On April 3, 2018, the only active region on the Sun was labeled as active region (AR) NOAA 12703. Figure 1a displays the line-of-sight magnetogram observed by SDO/HMI at 03:49:15 UT, where the leading and following sunspots exhibit negative and positive polarities, respectively. GONG H$\alpha$ observations at 03:50:50 UT (see Figure 1b) reveal a single solar filament located near the solar equator and near the disk center. Figures 1c-1f present subimages of the full-disk GONG H$\alpha$, 304 \AA, 171 \AA, and 131 \AA\ images, with magnetic fields overlaid. In Figures 1c-1f, green and blue contours indicate positive and negative magnetic fields, respectively, with contour levels of $\pm$100 G and $\pm$400 G. The magnetic fields predominantly show positive polarity on the eastern side and negative polarity on the western side of the filament. The filament crosses the center of AR NOAA 12703 and lies along the polarity inversion line (PIL). One end of the filament is anchored in the leading negative polarity, while the other end is connected to a quiet Sun location distinct from the active region (see Figures 1c and 1d). According to Engvold's (1998) classification, this filament is categorized as an intermediate filament. SDO/EUV observations indicate a filament channel to the south of the AR, where the filament resides, as pointed out by the white arrows in Figures 1e and 1f. Three observed characteristics support the existence of the filament channel: a relatively dark area between opposing magnetic polarities (see Figure 1c), the formation of the filament itself, and the alignment of thin threads/fibrils along the polarity boundary (see the yellow arrows in Figure 1e). In addition, the fibrils with anti-parallel alignment on either side of the PIL and the filament were also observed by the NVST in H$\alpha$ line center images on April 5.

\subsection{Cool material injection from a B-class flare}
The appearance of the AR NOAA 12703 in the chromosphere exhibits two bright patches near the northern end of the filament. From a series of H$\alpha$ images observed by GONG, the material motion in the filament can be seen clearly (see the supplementary movie 1). From 00:00 UT to 03:20 UT, the material with low temperature is found to be injected continuously into the filament from the AR NOAA 12703 (see Figure 2a-2c). During its evolution, a barb appears in the H$\alpha$ observation at 03:36:50 UT (X: -390, Y: -170) (see Figure 2c), where the magnetic structure of the filament becomes wide with obvious twisted magnetic field lines displayed in the following figures. According to the soft X-ray flux observation (see the yellow curve line in Figure 3), a small flare (B2.1) occurring in this AR started at about 03:56 UT, peaked at 04:05 UT, and ended at 04:15 UT. The obvious brightening in H$\alpha$ images can be seen in Figure 2e. After the occurrence of the flare, the material of the filament is pushed from the AR to the southern end of the filament. The cool material of the filament shown in Figure 2e-2f can be seen to move toward the southern part of the filament. At 04:26:50 UT, only a small amount of cool material is visible at the southern end of the filament. Some of the filament material is heated and disappeared in the H$\alpha$ line center observations as it moved southward. The following Figures 3 and 4 and supplementary movies 1 and 2 also demonstrate that the material is moving southward.

Due to the limitation of the field of view of the NVST, only the northern part of this filament was observed. The H$\alpha$ line center images and Doppler maps constructed using H$\alpha$ off-band images ($\pm$ 0.5 \AA) are presented in Figures 2g-2i and 2j-2l, respectively. The field of view of the NVST is indicated by the black box in Figure 2d. A series of H$\alpha$ center images reveals that the northern part of the filament has a curved shape and crosses the active region (see Figure 2g). Following the onset of the B-class flare, brightening occurs around the polarity inversion line (PIL) in the active region. Subsequently, several flare ribbons appear on both sides of the PIL (see Figure 2h). A bright patch appeared in the eastern part of the active region during the B-class flare. We reviewed the evolution of the line-of-sight magnetic fields observed by SDO/HMI and found no significant evidence of magnetic emergence or cancellation in that area. Therefore, we suggest that this bright patch was not caused by a jet-like eruption but may be related to magnetic reconnection during the B-class flare, which differs from the simulation by Luna \& Moreno-Insertis (2021). Before and after the B-class flare, the Doppler maps show noticeable opposite Doppler shifts on either side of the filament spine (see black and blue arrows in Figures 2j-2l). Note that the black and the blue arrows indicate the red-shifted and the blue-shifted signals, respectively. These observations indicate a continuous rolling motion in this region of the filament. During the B-class flare, the northern part of the filament has not changed and retained its original state.

To determine the velocity of material motion within the filament, a time-distance diagram was created along the filament's spine using GONG H$\alpha$ images taken between 00:00 UT and 06:00 UT on April 3, 2018. The curved cutout marked by the yellow lines in Figure 2b serves as the basis for the time-distance diagram shown in Figure 3. The yellow curve represents the profile of the GOES soft X-ray flux in the 1-8 \AA\ range. The red lines indicate intermittent material injections along the filament's spine prior to the B-class flare. Four material injections are observed moving from the northern end of the filament to the southern end, with velocities ranging from approximately 20 km $s^{-1}$ to 40 km $s^{-1}$. When the B-class flare occurred in AR NOAA 12703, the filament's material is driven to move from its northern end to the southern part, with a velocity of about 43 km s$^{-1}$ (see the cyan line in Figure 3). Note that Figure 3 is oriented with north at the bottom and south at the top.

The observations from the Optical and Near-infrared Solar Eruption Tracer (ONSET) captured the entire process of flare material injection. Figure 4 presents H$\alpha$ center images (Figures 4a1-4a4), H$\alpha$ +0.5 images (Figures 4b1-4b4), and H$\alpha$ -0.5 images (Figures 4c1-4c4) from 03:56 UT to 04:10 UT (see Supplementary Movie 2). It is evident that cool material is pushed from the active region to the southern end of the filament along its spine after the B-class flare occurs (see Figures 4a1-4a4). The black arrows in Figures 4a1-4a4 points to the southern end of the moving cool material. After the onset of the flare, the material of the filament is pushed toward the southern part of the filament. One can see that the filament become short with time and only the southern part of the filament was visible (see Figures 4a1-4a4). Additionally, red and blue shifts are observed along the filament spine (see Figures 4b1-4b4 and 4c1-4c4). The black arrows in Figures 4b1–4b4 indicate redshifted signals, which shifted slightly toward the northwest. In contrast, the black arrows in Figures 4c1–4c4 represent blueshifted signals, which moved southward. The reconstructed Doppler maps in Figures 4d1–4d4 confirm the southward movement of the blueshifted signals. Note that the black and the blue arrows denote the redshifted and blueshifted signals, respectively. Both the northern and southern parts of the filament exhibit simultaneous redshifted and blueshifted signals. This suggests that some injected material moved toward the south, while others were directed back toward the north along the filament spine.

\subsection{Fine structure of the filament traced by the hot material injection}
\subsubsection{The hot material injection in SDO/EUV observations}
In the H$\alpha$ images, cool material is clearly observed being pushed from the northern part of the filament to the southern part after the occurrence of the B-class flare. In contrast, the EUV images captured by SDO/AIA show only hot material moving from the northern to the southern part of the filament. Initially, the structure of the filament appears very narrow (see Figures 5a-5c). Subsequently, the filament's structure expands, revealing intertwining helical magnetic field lines (see Figures 5d-5i). When the hot material reach the southern end of the filament, a distorted magnetic structure becomes clearly visible (see Figures 5j-5l). The process of hot material injection observed in the SDO 304 \AA\ images can be viewed in Supplementary Movie 3.

Figures 6a1-6d1 display composite images created using observations from three different wavelengths (304 \AA, 171 \AA, and 211 \AA) (see Supplementary Movie 4). In these images, red, green, and blue colors correspond to the 211 \AA, 171 \AA, and 304 \AA\ observations, respectively. Following the injection of hot material after the B-class flare, the intertwined magnetic structure can be gradually traced. The observations at four different moments (03:58:35 UT, 04:08:11 UT, 04:12:47 UT, and 04:28:59 UT) are presented in Figures 6a1-6d1. In the corona, the magnetic pressure significantly exceeds the gas pressure, causing plasma movement to predominantly follow the magnetic field lines. Thus, the movement of the plasma allows for approximate tracing of the magnetic field lines.

Figures 6a2-d2 display composite images with traced magnetic field lines superimposed. The white line indicates the location of the filament channel as observed in the EUV images. The yellow and pink lines in Figures 6a-6c represent helical magnetic field lines traced by the moving hot material observed in the 171 \AA\ images. It is important to note that the yellow and pink lines in Figure 6 correspond to the same magnetic field lines observed at different times.

At the onset of the flare at 03:58:35 UT, hot material is injected into the filament, with the yellow and pink lines outlining the magnetic field lines traced by this injection (see Figure 6a2). By 04:08:11 UT, the two sets of magnetic field lines are intertwined (see Figure 6b2). Four minutes later, the structure marked by the pink line merges with the filament spine indicated by the red line (see Figure 6c2). Additionally, a relatively strong twisted magnetic structure is located at the center of the filament (also see white arrows in Figures 5g-5i). The tightly wrapped magnetic field lines enclose the filament material at its core (see Supplementary Movie 5). The red line outlines the filament spine with a twisted magnetic structure in Figure 6c2.

Based on the evolution of the traced magnetic field lines, the two golden lines in Figure 6d2 delineate the main body of the filament's magnetic structure. The aquamarine lines in Figures 6c2 and 6d2 indicate field lines lying above the filament. When the injected flare material reaches the southern end of the filament, a kinked structure appears (see Figures 6d1, 6d2, and 5j-5l). This resembles the writhe of solar filaments and flux ropes during their eruptions, as noted in previous studies (Ji et al. 2003; Kliem et al. 2004; T{\"o}r{\"o}k, \& Kliem 2005; Liu \& Alexander 2009). It is noteworthy that the filament does not ultimately erupt.

\subsubsection{Comparison between SDO/EUV and H$\alpha$ observations}
To compare the appearance of the filament in H$\alpha$ and EUV images, composite images created from both types of observations are shown in Figure 7. In these images, green represents the 171 \AA\ observation, while red represents the H$\alpha$ observation. As is well known, the filament appears more extended in the EUV wavelengths than in the H$\alpha$ band due to the absorption of background EUV radiation by the Lyman continuum. The maximum width of the filament in the H$\alpha$ images is about 20 arcseconds (before the material injection from the B-class flare), whereas in the 171 \AA\ images, it reaches about 60 arcseconds.

The cool, dense filament material is enveloped by strong twisted magnetic field lines in the filament's core (see Figure 7a). Consequently, H$\alpha$ observations primarily reveal the core of the filament, which contains low-temperature, high-density material, while EUV observations display the surrounding magnetic structure. Additionally, the moving filament material stops at the base of the kinked magnetic structure (see Figure 7b).

Before the hot material injection from the B-class flare, these magnetic structures were not visible in any UV or EUV observations. After the injection, the magnetic field lines can be gradually traced. These observations suggest that the temperature of the traced magnetic structures is lower than that of the EUV line (304 \AA, 5$\times$ $10^{4}$ K). Thus, the magnetic structures revealed by the hot flare material injection may correspond to the prominence-corona transition region (PCTR) (Labrosse et al. 2010), which has not been observed previously.

\subsubsection{The fine-scale magnetic structures revealed in 171  \AA\ observations}
To illustrate the magnetic structures of the filament in detail, Figures 8a and 8c present 171 \AA\ images observed at 04:11:34 UT and 171 \AA\ running difference images (04:11:34 UT - 04:11:10 UT), respectively. Due to the hot material injection, previously invisible magnetic structures became visible, particularly as thin thread-like structures in the SDO/EUV observations. It is clear that the main body of the filament has a complex structure, as highlighted by the blue boxes in Figures 8a and 8c.

To enhance visibility of this structure, Figures 8b and 8d show amplified sections of the filament from the blue boxes in Figures 8a and 8c. Several magnetic structures are traced and marked by red and pink lines in Figures 8b and 8d. The red and pink dotted lines indicate the mutually wrapped and twisted magnetic structures that enclose and compose the filament's magnetic configuration.

\subsubsection{DEM analysis}
During the material injection, the magnetic structures of the filament sequentially became bright, moving from the northern part to the southern part. The left panels of Figure 9 display emission measure (EM) maps (Figures 9a-9c), while the right panels show temperature maps calculated using six wavelength images obtained from SDO/AIA (Figures 9d-9f) at four different moments. After the hot material injection, the evolution of the EM and temperature maps reveals the profile of the filament.

The magnetic structures of the filament gradually heated up, resulting in brightening. The structures in the northern part were the first to heat and subsequently extended to the southern part of the filament. The maximum temperature of the traced magnetic loops reaches approximately 10 MK. It is important to note that these magnetic structures were not visible in any EUV observations prior to the injection of hot flare material.

\subsubsection{Reconstructed filament magnetic structures}
To obtain the magnetic structures of the filament, we combined images of its three parts taken at different moments using EUV images from the SDO. In other words, images captured at different times were composited into a single image. Based on the evolution of material injection, we selected three observational moments to reconstruct these composite images. The reconstructed images are presented in Figure 10. These three selected moments encompass the entire process of material motion following the B-class flare. The colors red, green, and blue represent the traced structures at 04:00 UT, 04:07 UT, and 04:26 UT, using 304 \AA, 171 \AA, 211 \AA, and 131 \AA\ images. The reconstructed images clearly reveal the complete magnetic structure of the filament, showcasing an intertwined magnetic structure enveloping it. Additionally, a kinked section of the filament is visible at its southern end.

To compare the magnetic structures of the filament traced by the hot material injection with different methods, we also employ a maximum fusion method to reconstruct the magnetic structures depicted in Figure 11. This reconstruction utilizes a sequence of 171 \AA\ images captured at four different times: 04:04:09 UT, 04:07:33 UT, 04:17:21 UT, and 04:27:33 UT. We extract the intensity values of each pixel from these images, and the maximum intensity values are then used to create a single composite image.

When compared to the magnetic structures shown in Figure 10, the overall magnetic structure of the filament in Figure 11 appears similar. However, the width of the filament in Figure 11 is notably greater than in Figure 10. Some finer details of the filament's structure are lost due to the overlapping features from the four different time points. The advantage of this method lies in its ability to capture the overall structure of the filament effectively.

\subsubsection{Filament magnetic structures seen from different observational angles}
This filament is simultaneously observed by the SDO and STEREO Ahead satellites. To compare observations from these different perspectives, we present the on-disk observations from SDO alongside the limb observations from STEREO. Figure 12 illustrates the evolution of the filament as observed by both satellites during the hot flare material injection. The first and second columns display the 171 \AA\ images and the 171 \AA\ running difference images at four different moments, while the third and fourth columns show the 304 \AA\ images and the 195 \AA\ images captured by STEREO, respectively. These observations reveal that the magnetic structure of the filament appears significantly different when viewed from different angles.

From the on-disk AIA observations, the twisted magnetic structures are clearly visible (see the first and second columns of Figure 12). The red lines in Figures 12a, 12e, and 12i denote the spine of the filament. The green and cyan lines represent two groups of intertwined magnetic field lines in Figure 12a, which are also evident in the running difference image in Figure 12b. Approximately 10 minutes later, at 04:16:10 UT, the structure marked by the green line merges with the main body of the filament and is not shown in Figure 12e. However, the magnetic field line indicated by the cyan line remains intact and continues to intertwine with the main body of the filament. A careful analysis of the magnetic structure's evolution in the 171 \AA\ observations reveals another set of magnetic loops, marked by the blue lines in Figures 12e and 12i. This structure is not very prominent in the running difference images shown in Figures 12f, 12j, and 12n, but a kinked structure at the southern end of the filament is clearly visible in all three images. These field lines lie along the body of the filament, which is also apparent in the 304 \AA\ observations in Figure 5. Furthermore, we observe that the filament consists of different segments of magnetic field lines associated with the magnetic polarities along its main body. As the injected material descends along these magnetic field lines, brightening occurs at their footpoints. The red circles in Figure 12m indicate the footpoints of the filament traced by the falling material.

At the solar limb, the northern part of the filament appears lower-lying, while the southern part extrudes from the Sun at the onset of the B-class flare (see Figures 12c and 12d). The white arrows in these figures indicate the rising structure of the filament following the hot flare material injection. At 04:16:15 UT, the body of the filament resembles a bridge (see Figures 12g and 12h), as marked by the white arrows, suggesting that the magnetic structure of the filament resembles a flux rope. Subsequently, the filament transforms into a wall-like structure after material is injected into its southern part (see Figures 12k, 12l, 12o, and 12p, as well as Supplementary Movie 6), as indicated by the white arrows. This change in emission structure results from material falling along the magnetic field lines that connect the magnetic polarities along the polarity inversion line. The emission structure in the 195 \AA\ images is not as pronounced (see the fourth column of Figures 12l and 12p). Notably, the kinked structure observed at the southern part of the filament in SDO/EUV observations is not visible in the limb observations. We suspect that this kinked structure may be obscured by other parts of the filament. 

\subsubsection{Cartoon model for the filament}
Considering the observations, we draw a schematic diagram to illustrate the magnetic structures of the intermediate filament (see Figure 13). The black line represents the axis of the filament. The gray tube represent the compact set of mutually wrapped magnetic fields encasing the cool, dense filament material. The red, orange, and dark red lines represent the twisted magnetic field lines enclosing the intermediate filament. The blue lines represent the magnetic field lines wrapping the intermediate filament and rooted in other footpoints (not the footpoints of the filament), which are corresponding to the yellow lines in Figure 6 and the cyan lines in Figure 12. The yellow and green patches represent the negative and positive magnetic fields, respectively. Note that we ignore the magnetic field lines overlying on the filament (the aquamarine lines in Figures 6c-6d and the blue lines in Figure 12).

\section{Conclusion and Discussions}
Using multi-wavelength observational data from the NVST, ONSET, SDO, and STEREO, we investigate the magnetic structures of a solar intermediate filament traced by the injected hot material from a B-class solar flare in AR NOAA 12703. We employ four EUV wavelengths at three different times and use a maximum fusion method to reconstruct the filament's complete magnetic structures, respectively. We also analyze the evolution of the emission measure and temperature of the filament during the hot material injection. Through analyzing these high-resolution H$\alpha$ images alongside multi-wavelength EUV images, we find that the filament consists of a compact set of mutually wrapped magnetic fields encasing the filament material, two distinct sets of magnetic field lines intertwined with the main filament, and a series of arched magnetic loops positioned along its length. Additionally, the northern footpoints of the helical structures are rooted in the same location, while their southern footpoints originate from different areas.

Some observations and theoretical modelling indicate that filaments are a multi-thermal object with a cool core and a hot shroud surrounding. The temperature of the shroud surrounding can reach coronal values (Heinzel et al. 2016). Due to the limited spatial resolution of current magnetic measurements, it is difficult to accurately determine the magnetic structure of solar filaments and prominences. Present observations primarily capture emissions from different plasmas within solar filaments. However, the magnetic structures enveloping the filament material are largely invisible before filament eruptions, whether in on-disk or limb UV (or EUV) observations.

In fact, the structures seen in EUV observations result from plasma emissions, not the actual magnetic structures. In the corona, the plasma beta (the ratio of gas dynamic pressure to magnetic pressure) is very low, and the plasma's electrical conductivity is very high. As a result, injected plasma flows along magnetic field lines rather than across them. Due to the frozen-in condition of the corona, the injected material in this event moves along the magnetic field lines. The trajectories of these injected material thus trace the magnetic field lines of the filament. Based on this, the magnetic structures of the filament can be revealed by following the path of hot material injected from a B-class flare in AR NOAA 12703. It is important to note that this filament experienced a failed eruption following the injection of hot flare material.

Moreover, a rolling motion was detected in the filament using Doppler maps constructed from H$\alpha$ off-band images observed by the NVST. Additionally, material from the filament is continuously injected into it from its northern end, as clearly seen in the GONG H$\alpha$ observations before the onset of the B-class flare (see supplementary movie 1 and the time-distance diagram in Figure 3). This supports the idea that the filament material originates from the lower atmosphere (Wang et al. 2018).

We also observed a brightening patch to the east of the active region when the B-class flare occurred. To investigate this further, we analyzed the evolution of line-of-sight magnetic fields using SDO/HMI observations but did not find any significant magnetic emergence or cancellation. This rules out the possibility of jets or other related activities occurring in that region. However, in the 171 \AA\ observations, post-flare loops appeared at 04:18:21 UT (see supplementary movie 5), located above the northern part of the filament. Based on these observations, we speculate that the B-class flare may have been triggered by the partial eruption of filament threads. 

Previous observations indicate that H$\alpha$ filaments appear narrower than their EUV counterparts due to Hydrogen Lyman continuum absorption (Heinzel et al. 2001). These observations also reveal that the filament in H$\alpha$ represents only the core of a larger magnetic structure and occupies about one-third of the large-scale magnetic structure, known as the EUV filament channel. This finding is slightly larger than the result of Zhou et al. (2020), who found that the cold material represented by filament threads occupies about 10-15\% of the filament channel. The compact, interwoven magnetic structures surrounding the cool, dense filament material enable it to remain suspended in the hot, tenuous solar corona.

This study also reveals that the filament material is enveloped by mutually wrapped magnetic structures. Additionally, a group of magnetic field lines intertwines with the main body of the filament. The northern footpoints of this group and the primary magnetic structures of the filament are rooted in the same location, while their southern footpoints are anchored in different places. These observed structures may confine the filament, preventing its eruption. The magnetic structures of the intermediate filament observed in this event confirm the model of filament magnetic structure proposed by Aulanier \& Schmieder (2002). The complex magnetic structures of solar filaments, particularly in quiescent filaments and prominences, require higher-resolution observations to fully understand their nature in the future.

Acknowledgments: We thank the anonymous referee very much for constructive suggestions and comments that significantly improved the original version of the manuscript. We would like to thank the NVST, SDO, STEREO, ONSET teams for high-cadence data support. This work is sponsored by the Strategic Priority Research Program of the Chinese Academy of Sciences, Grant No. XDB0560000, the National Science Foundation of China( NSFC) under the numbers 12325303, 11973084, 12003064, 12203097, Yunnan Key Laboratory of Solar Physics and Space Science under the number 202205AG070009, Yunnan Fundamental Research Projects under the numbers 202301AT070347, 202301AT070349. 


 \begin{figure*}
  \centering
   \includegraphics[width=15cm]{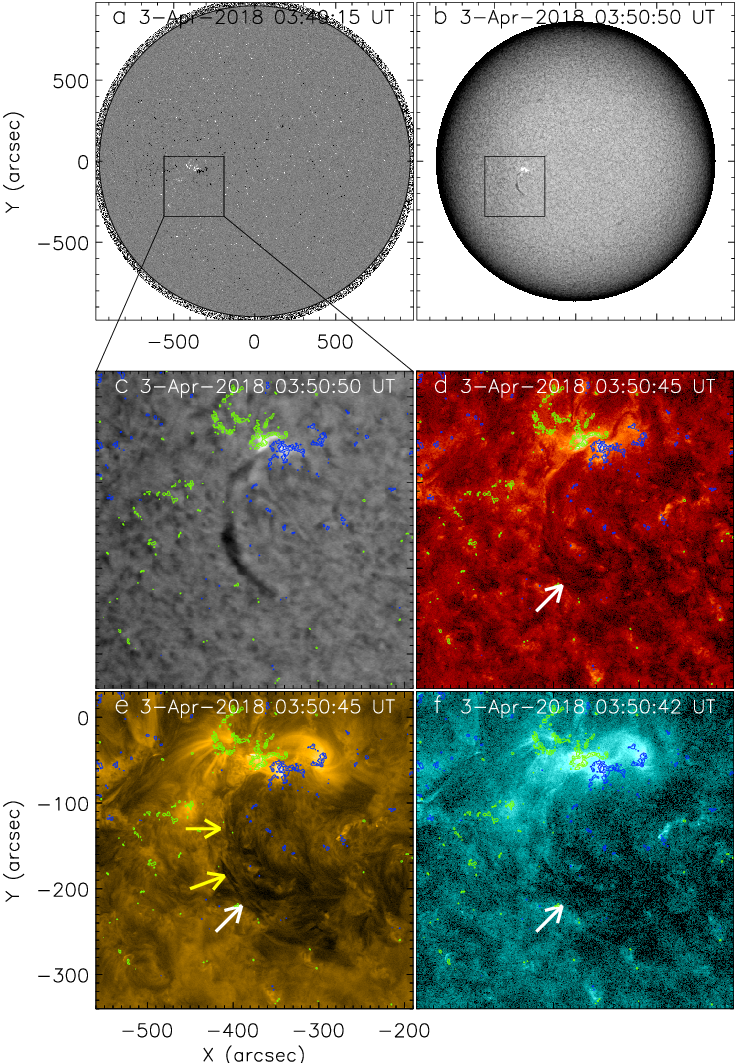}
\caption{{\bf Appearance of an intermediate filament.} (a): A full disk line-of-sight magnetogram observed by SDO/HMI at 03:49:15 UT on 2018 April 3. The white (black) patches indicate the positive (negative) magnetic field. (b): A full disk H$\alpha$ image observed by GONG at 03:50:50 UT. (c-f): sub-images of the full disk images observed by GONG and SDO/AIA. H$\alpha$ (Figure 1c), 304 \AA\ image (Figure 1d),  171 \AA\ image (Figure 1e), 131 \AA\ image (Figure 1f), respectively. The green and the blue contours in the Figures 1c-1f denote the positive and the negative magnetic fields. The levels of the contours are $\pm$ 100 G and $\pm$ 400 G, respectively. The white arrows in Figures 1c-1f denote the filament channel seen in EUV images.
}
     \label{fig1}
   \end{figure*}
 
   \begin{figure*}
  \centering
   \includegraphics[width=15cm]{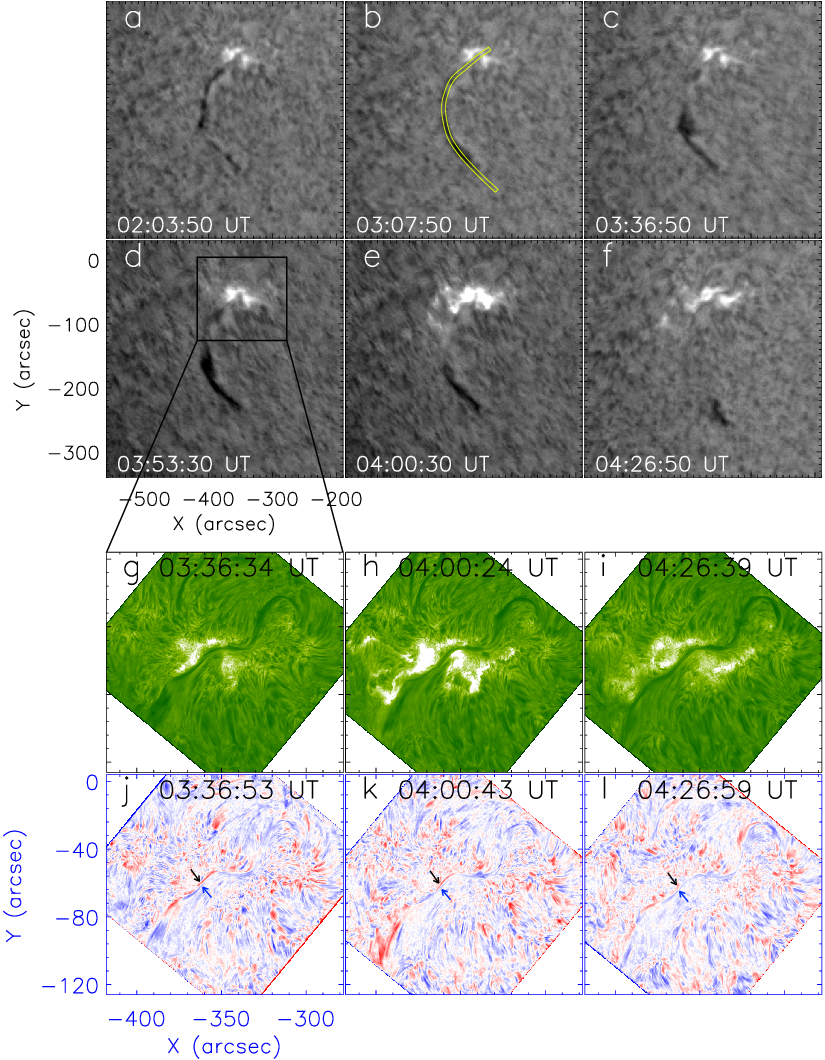}
\caption{{\bf Motion of the filament material and the occurrence of a B-class flare.} (a-f) GONG H$\alpha$ images showing the material motion from the northern part to the southern part along the spine of the filament and the occurrence of a B-class flare in AR NOAA 12703. (g-i) NVST H$\alpha$ center images showing the evolution of the northern part of the filament and the occurrence of a B-class flare. (j-l) Doppler maps constructed by using H$\alpha$ off-band images ($\pm$ 0.5 \AA). The red and the blue colors show the red and the blue shifts, respectively. The black and the blue arrows indicate the red-shifted and the blue-shifted signals, respectively. An animation of the GONG H$\alpha$ images is available. Its duration is 8 s, covering from 01:43 UT to 06:00 UT on 2018 April 3.
}
       \label{fig1}
   \end{figure*}
   
        \begin{figure*}
  \centering
   \includegraphics[width=15cm]{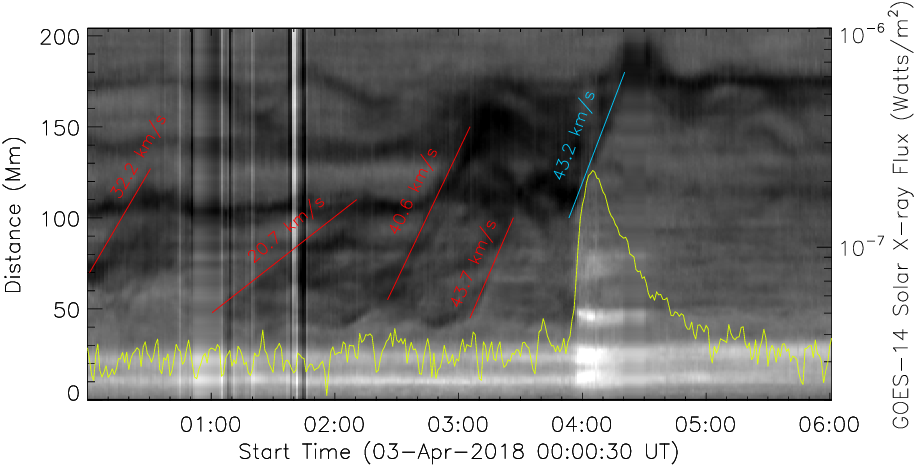}
\caption{{\bf A time-distance diagram along the spine of the filament made by using GONG H$\alpha$ images.} The red lines denote the intermittent material injection along the curved cutout marked by the yellow lines in Figure 2b. The cyan line denotes the material motion driven by the B-class flare. Note that this figure is oriented with the north at the bottom and the south at the top. The yellow curve line shows the profile of the GOES soft X-ray flux in 1 - 8 \AA.}
       \label{fig3}
   \end{figure*} 

     \begin{figure*}
  \centering
   \includegraphics[width=15cm]{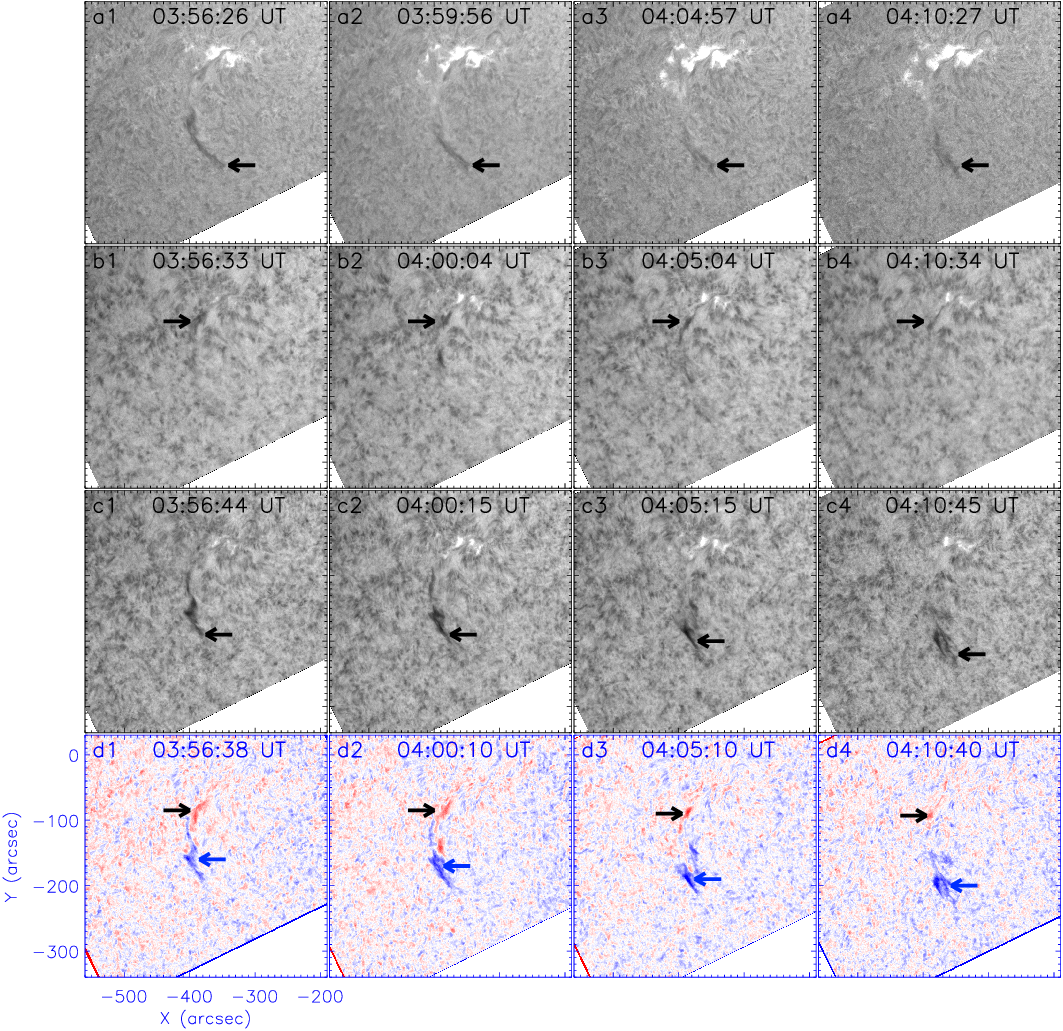}
\caption{{\bf Evolution of the filament and Doppler shift during the B-class flare from 03:56 UT to 04:10 UT observed by the ONSET.} (a1-a4): H$\alpha$ center image; (b1-b4): H$\alpha$ +0.5 images; (c1-c4): H$\alpha$ -0.5 images. (d1-d4): Doppler maps constructed by using H$\alpha$ off-band ($\pm$ 0.5 \AA) images. Note that The black arrows in Figures 4a1-4a4, 4b1-4b4, and 4c1-4c4 denote the southern end of the filament, the redshifted, the blueshifted signals, respectively. The black and the blue arrows denote the redshifted and blueshifted signals, respectively. An animation of the ONSET H$\alpha$ images is available. Its duration is 9 s, covering from 03:56 UT to 09:19 UT on April 3.}
       \label{fig3}
   \end{figure*} 
   
        \begin{figure*}
  \centering
   \includegraphics[width=15cm]{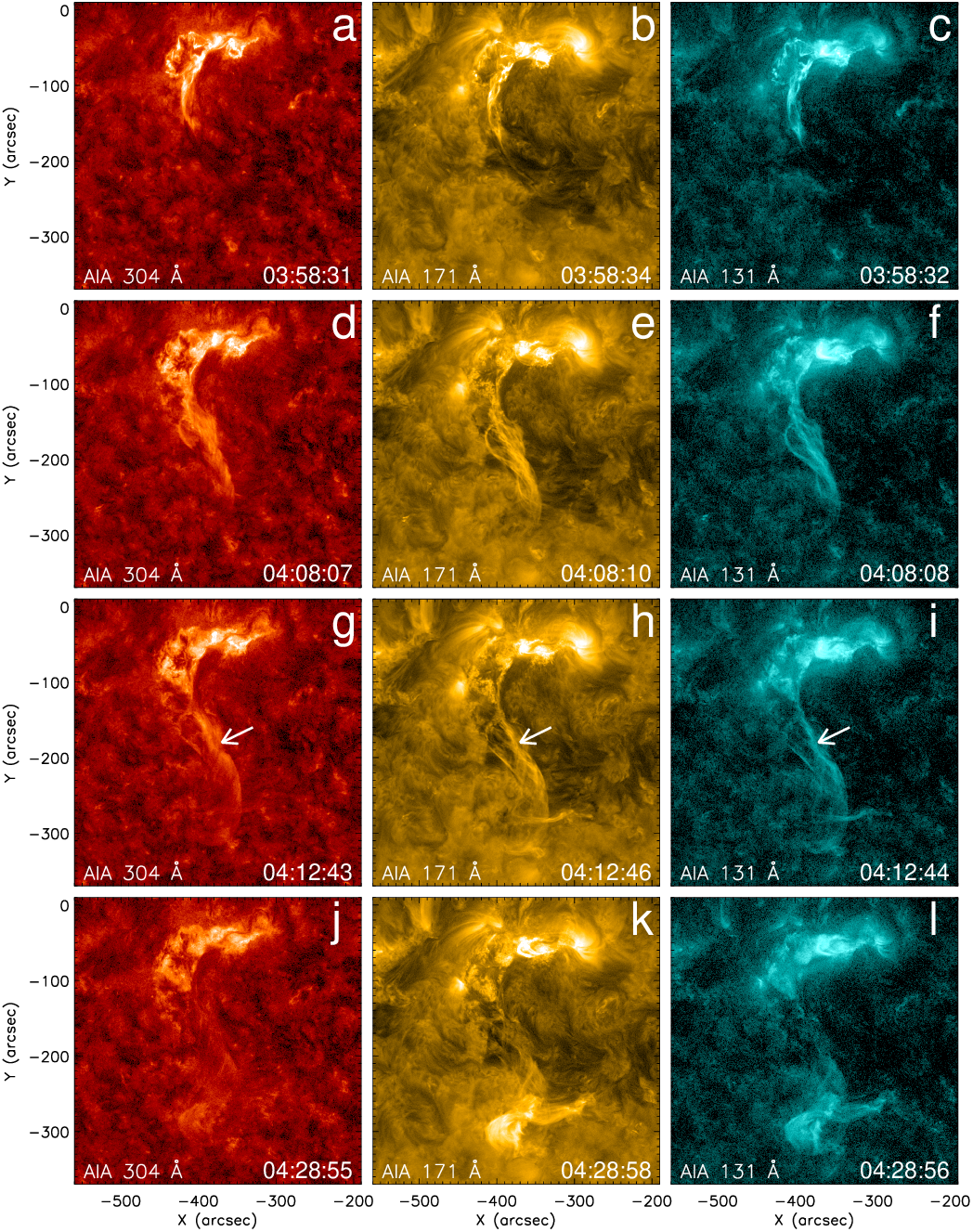}
\caption{{\bf Magnetic structures of the filament traced by the injected hot flare material.} (a-l) AIA EUV images at 304~\AA\ (a, d, g, and e), 171~\AA\ (b, e, h ,and k), and 131~\AA\ (c, f, i, and l) channels, in which the interwinding magnetic structures are traced by the hot material motion after the occurrence of the B-class flare. An animation of the 304~\AA\ images is available. Its duration is 25 s, covering from 03:29 UT to 05:29 UT on April 3.}
       \label{fig4}
   \end{figure*} 
  
    \begin{figure*}
  \centering
   \includegraphics[width=11cm]{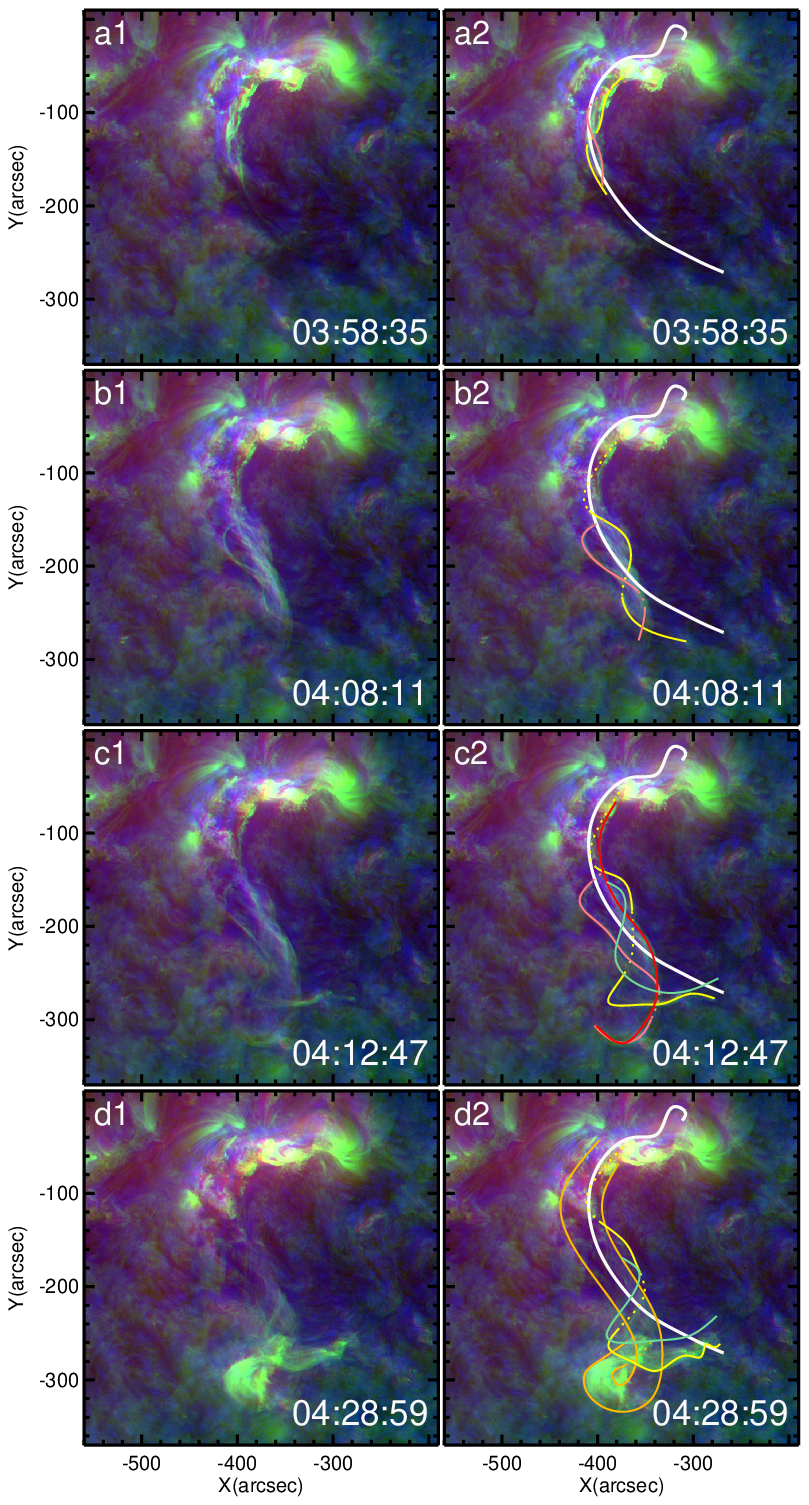}
\caption{{\bf Composite images showing the magnetic structure of the filament.} (a-d): Four composite images by using three different wavelengths (304 \AA, 171 \AA, and 211 \AA\ images) at 03:58:35 UT, 04:08:11 UT, 04:12:47 UT, and 04:28:59 UT, respectively. The white line denotes the filament channel. The yellow and the pink lines denote the interwinding magnetic field lines in Figures 6a, 6b, and 6c. The aquamarine lines in Figures 6c and 6d denote the magnetic loops lying on the main body of the filament. The outlined structure by the golden line is the main body of the filament. An animation of the composite images is available. Its duration is 25 s, covering from 03:29 UT to 05:29 UT on 2018 April 3.}
       \label{fig2}
   \end{figure*}

     \begin{figure}
  \centering
   \includegraphics[width=16cm]{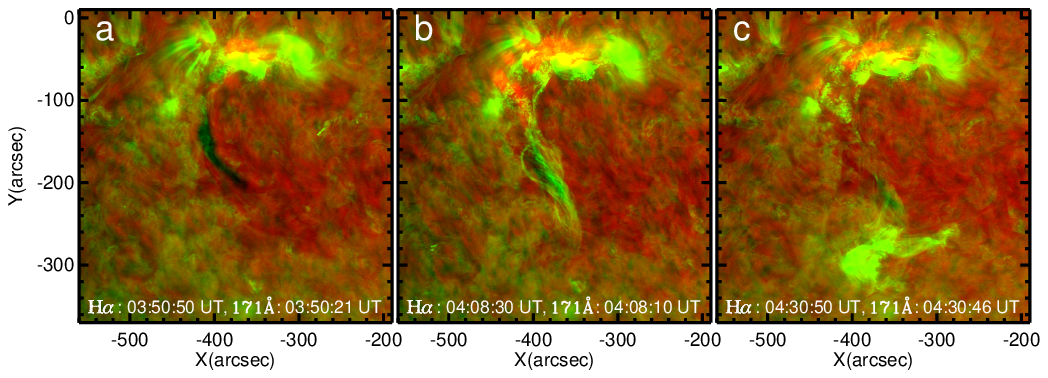}
\caption{{\bf Composite images of H$\alpha$ image and EUV image.} (a) The composite images of H$\alpha$ and EUV images at 04:08:10 UT. (b) The composite images of H$\alpha$ and EUV images at 04:30:46 UT. The green color represents the 171 \AA\ observation and the red color represents the $H\alpha$ observation.}
       \label{fig5}
   \end{figure} 
   
    \begin{figure*}
  \centering
   \includegraphics[width=13cm]{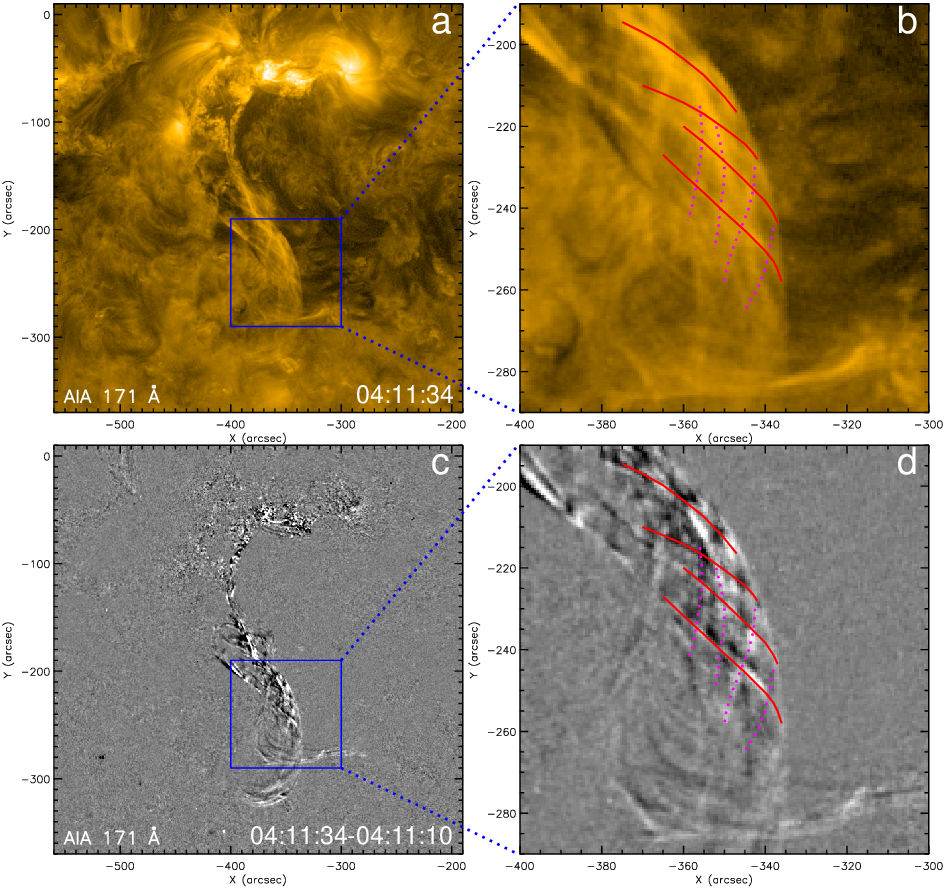}
\caption{{\bf 171 \AA\ and the corresponding running difference images.} A 171 \AA\ and a 171 \AA\ running difference images are shown in the first column, respectively. The amplified images in the blue boxes in Figures 8a and 8c are shown in Figures 8b and 8d, respectively. The solid red lines and the dashed pink lines show the compact mutually wrapped magnetic structures that are enwrapping the cool and dense filament material. {\bf Note that the red field lines are located at the top of the filament structure and the pink ones are located at the bottom.} An animation of the 171~\AA\ images is available. Its duration is 25 s, covering from 03:29 UT to 05:29 UT on April 3.}
       \label{fig9}
   \end{figure*}
   
   \begin{figure*}
  \centering
   \includegraphics[width=10cm]{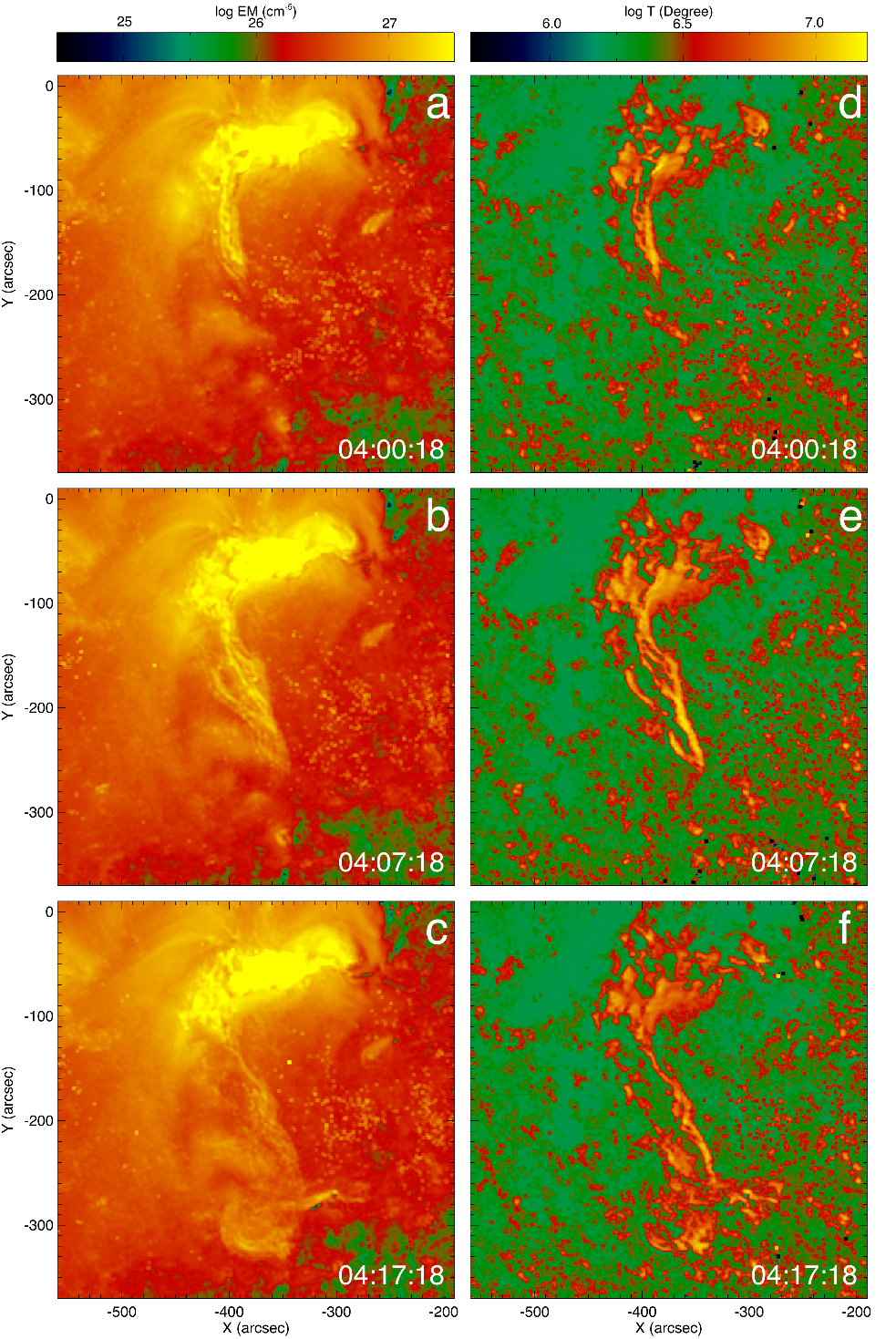}
\caption{{\bf Reconstructed EM map and temperature map.} (a-c): Reconstructed EM maps by using 171 \AA, 211 \AA, 193 \AA, 131 \AA, 335 \AA, and 94 \AA\ images at 04:00:18 UT, 04:07:18 UT, and 04:17:18 UT, respectively. The different colors indicate the different values of EM. (d-f): Calculated temperature maps corresponding the EM maps.}
       \label{Supplementary Figure 3}
   \end{figure*}

  \begin{figure*}
  \centering
   \includegraphics[width=13cm]{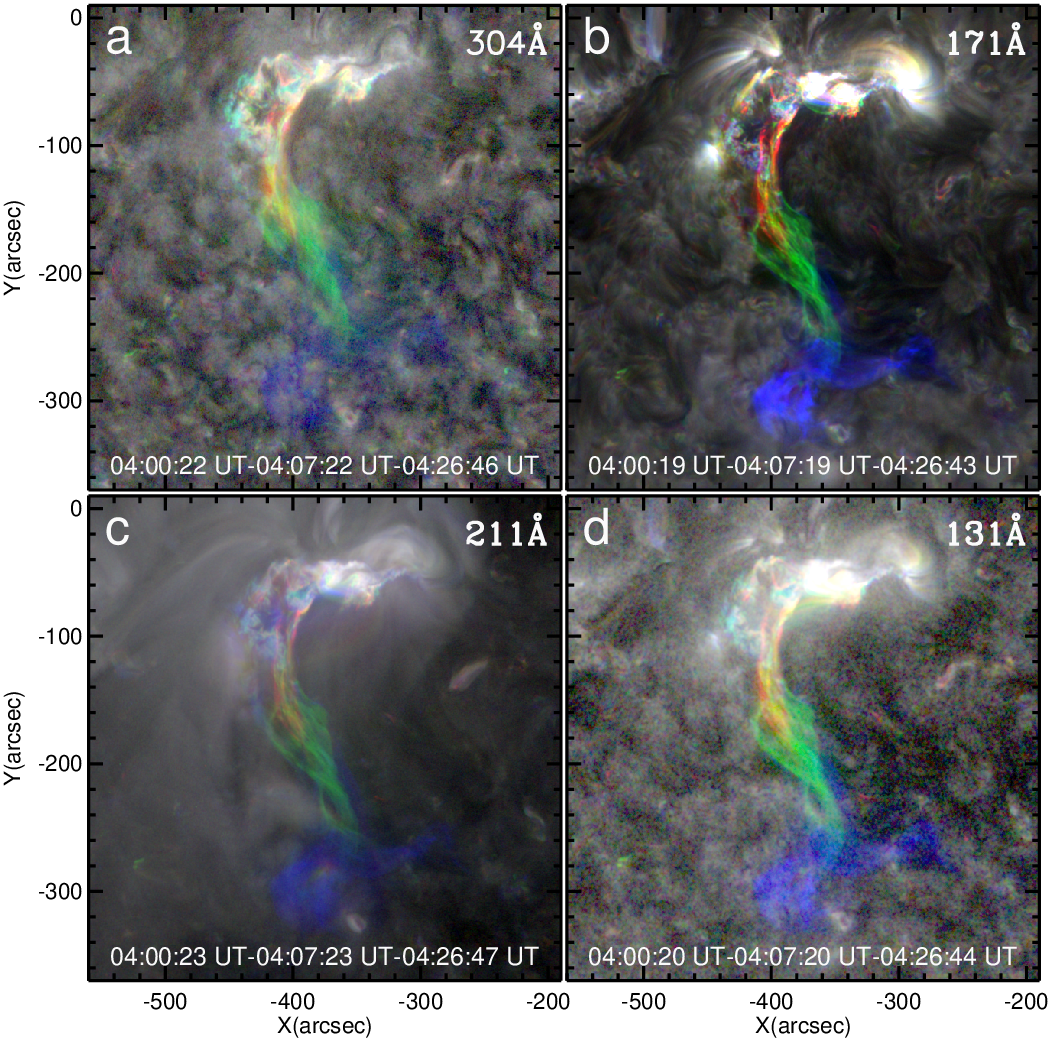}
\caption{{\bf Reconstructed filament magnetic structure by using the EUV images at three different times.} (a): Reconstructed filament magnetic structure by using 304 \AA\ images at three different moments. (b): Reconstructed filament magnetic structure by using 171 \AA\ images. (c): Reconstructed filament magnetic structure by using 211 \AA\ images. (d): Reconstructed filament magnetic structure by using 131 \AA\ images. Note that the red, the green, and the blue colors denote the traced structure at 04:00 UT, 04:07 UT, and 04:26 UT by using 304 \AA, 171 \AA, 211 \AA, and 131 \AA\ images, respectively.}
       \label{fig3}
   \end{figure*}.
  
  \begin{figure*}
  \centering
   \includegraphics[width=13cm]{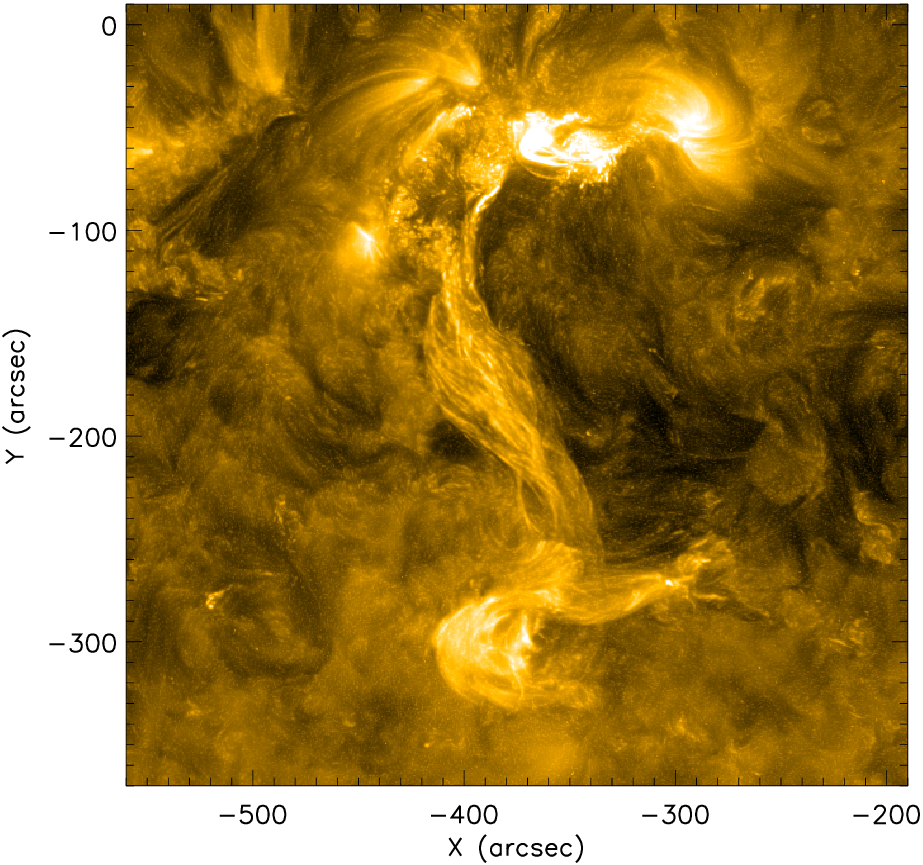}
\caption{{\bf Reconstructed magnetic structure of the filament by using the maximum fusion method.} The magnetic structure of the filament is  reconstructed by using 171 \AA\ wavelength images at four different moments (04:04:09 UT, 04:07:33 UT, 04:17:21 UT, and 04:27:33 UT). }
       \label{fig9}
   \end{figure*}

          \begin{figure*}
  \centering
   \includegraphics[width=17cm]{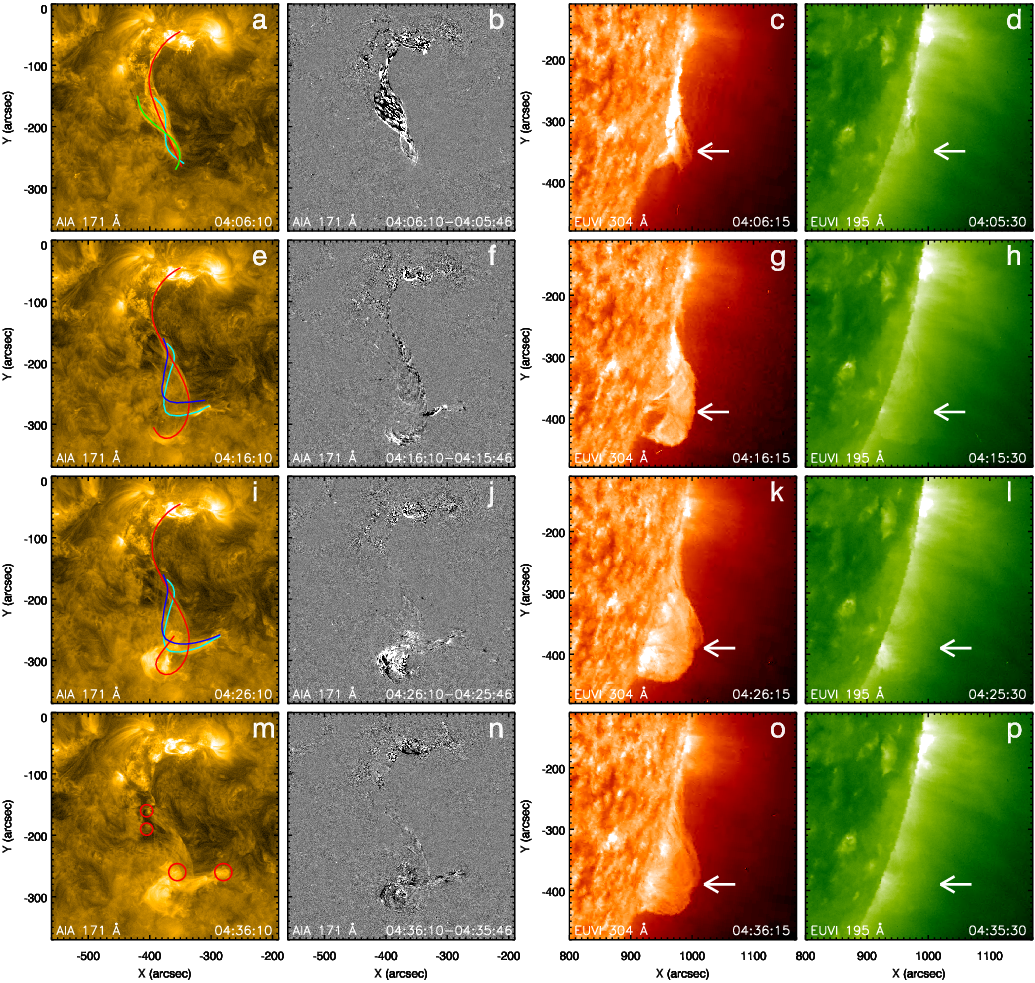}
\caption{{\bf The filament seen from two different viewing angles.} The first column and the second column show the 171 \AA\ images and the 171 \AA\ running difference images. The third column and the fourth column show 304 \AA\ images and 195 \AA\ image observed by the STEREO. The red line indicates the spine of the filament. The green and the cyan lines denote the two groups of the intertwined magnetic field lines in Figs. 12a and 12b. The cyan lines outline the field lines around the main body the filament and the blue lines outline a group of magnetic loops overlying the filament in Figs. 12e-12j. The red circles in Fig. 12m mark several foot-points of the filament traced by the falling material. The white arrows denote the filament when it is seen at the solar limb by the STEREO. An animation of the STEREO 304~\AA\ images is available. Its duration is 2 s, covering from 03:01 UT to 04:56 UT on April 3.}
       \label{fig9}
   \end{figure*}
   
             \begin{figure*}
  \centering
   \includegraphics[width=13cm]{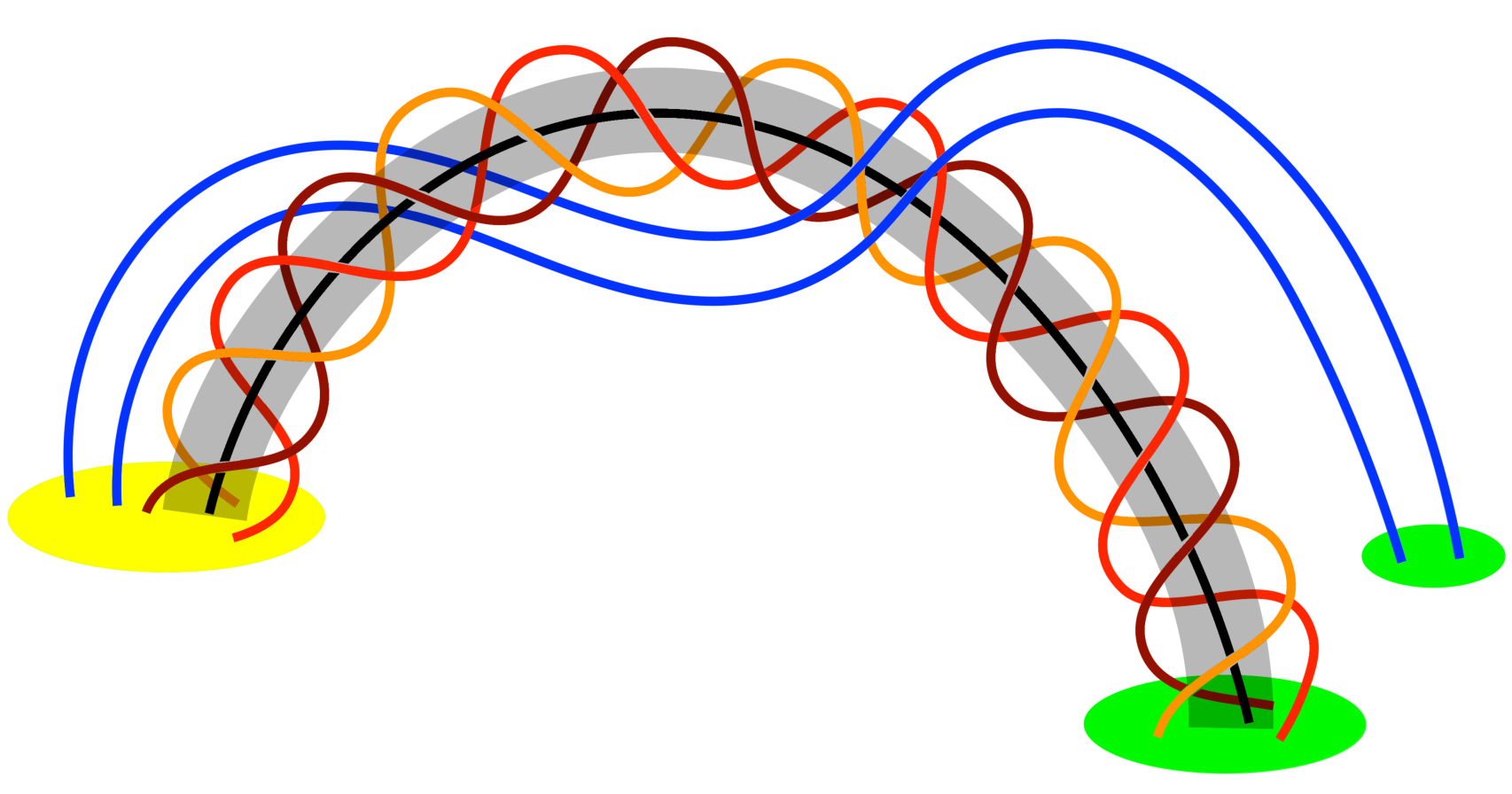}
\caption{{\bf Schematic diagram illustrating the magnetic structures of the intermediate filament.}  The black line represents the axis of the intermediate filament. The gray tube represent the compact set of mutually wrapped magnetic fields encasing the cool and dense filament material. The red, orange, and dark red lines represent the surrounding twisted magnetic field lines enclosing the intermediate filament. The blue lines represent the magnetic field lines wrapping the intermediate filament and rooted in other footpoints (not the footpoints of the filament). The yellow and green patches represent the negative and positive magnetic fields, respectively.}
       \label{fig9}
   \end{figure*}

\end{CJK*}
\end{document}